\documentstyle[12pt]{article}
\begin{document}
 
\title {\bf Transition between Two Oscillation Modes}
\author{R. L\'{o}pez-Ruiz and Y. Pomeau  \\                                  
{\small L.P.S., Ecole Normale Sup\'{e}rieure} \\
{\small 24 rue Lhomond, 75231 Paris Cedex 05 (France)} 
\date{ }}

\maketitle
\baselineskip 8mm

\begin{center} {\bf Abstract} \end{center}
A model for the symmetric coupling of two self-oscillators
is presented. The nonlinearities cause the system to vibrate in two modes
of different symmetries.
The transition between these two regimes of oscillation can
occur by two different scenarios. 
This might model the release 
of vortices behind circular cylinders with a possible transition
from a symmetric to an antisymmetric B\'{e}nard-von Karman vortex street. \par
{\small {\bf PACS number(s)}: 47.20.+k, 05.45.+b, 05.90.+m}

\newpage

Nowadays the understanding of self-oscillators is fairly
complete thanks to the theory of bifurcation and of normal forms.
A familiar model for this is the van der Pol system
\cite{vanderpol} that displays
a wide range of behavior, from weakly nonlinear to strongly
nonlinear relaxation oscillations, making it a good model for
many practical situations. However there are physical situations 
characterized by spontaneous self-oscillations with certain basic 
features that are absent from the "generic" van der Pol system.
Take for instance the B\'{e}nard-von Karman vortex street
in the wake of a cylinder. Its phenomenology is approximately as
follows \cite{mathis,noack,leweke}:
the velocity field remains two-dimensional for
Reynolds number ($Re$) less than $160$ (creeping flow for $Re<4$ ;
recirculation zone with two steady symmetric eddies attached behind the cylinder
for $4<Re<45$ ; instability at $Re\simeq  45$ at wich these eddies 
are released alternatively to form a double row of opposite
sign vortices, the B\'{e}nard-von Karman vortex street) and for
$Re>160$ three-dimensional and irregular fluctuations are superimposed
on the dominant periodic vortex shedding. \newline
It is tempting to say that the periodic vortex shedding
provides a classical example of the Poincare-Andronov 
bifurcation to a limit cycle. However, one fundamental
ingredient would be missing if one insisted in describing these oscillations
by the van der Pol equation: no equivalent of the symmetry of the
system would be present in this mathematical description.
That is, this mathematical picture would make no difference
between a symmetric and an antisymmetric release of vortices,
both would be fairly described by the same van der Pol equation,
although they are clearly physically different. \newline
We propose here to implement the major symmetries of the 
B\'{e}nard-von Karman oscillations by assuming that they result
from the symmetric coupling of two identical oscillators,
each one responsible of the periodic release  of vortices
on one side of the cylinder. The interest of this approach
is that it shows two possible stable oscillating states:
one symmetric, one antisymmetric, depending
on the value of some coupling parameter.
By varying  continously the coupling, it is possible
to monitor the transition between these two regimes,
something that is beyond an approach using a single van der Pol 
equation. \par
A dynamical model for representing these properties is presented and
some informations about the transitions  between 
the two oscillation modes are obtained. We make no attempt to relate
our model to the fluid mechanical equations.
The symmetry properties of the system are used as a basic ingredient,as well as the
fact that it operates in a stable way in an oscillating mode.

Let us assume that there is an oscillator (the 'vortex' emitter) on each side
{\it (side$_1$,side$_2$)} of the cylinder \cite{pomeau1} and let 
$(x_1,x_2)$ be their amplitudes of oscillation. If $(x_1(t),x_2(t))$ is
a possible dynamics, $(x_2(t),x_1(t))$ is also realizable
by symmetry. If a vortex is emitted from {\it side$_i$} whenever $x_i(t)$
reaches a maximum then the symmetrical and antisymmetrical vortex streets (Fig. 1)
appear as two oscillation modes, one with the two oscillators in phase and
the other one with the two oscillators out of phase.
The symmetric mode $\Theta_1$ verifies $x_1(t)=x_2(t)$ and the
antisymmetric one $\Theta_2$ verifies $x_1(t)=x_2(t+T/2)$ with $T$ 
the period. \newline
The simplest model representing these properties is a system of 
two coupled harmonic oscillators with a small coupling $\beta$:
\begin{eqnarray}
\ddot{x}_1+x_1+\beta x_2 & = & 0 \nonumber \\
\ddot{x}_2+x_2+\beta x_1 & = & 0
\end{eqnarray}
The normal modes verify:
$\ddot{\Theta}_i+(1-(-1)^i\beta)\Theta_i=0$ $(i=1,2)$,
with $\Theta_1=x_1+x_2$ and $\Theta_2=x_1-x_2$, each one with frequencies
$\omega_1=1+\beta$ and $\omega_2=1-\beta$. But this model is Hamiltonian
and it is not useful to describe self-oscillations (i.e. oscillations
resulting from balance between energy input and dissipation).
It does not present any Poincar\'{e}-Andronov bifurcation,
although it was shown experimentally that the vortex shedding
behind a cylinder results from that type of
bifurcation \cite{mathis}.
To remedy this, we can introduce, as in the van der Pol 
system, a nontrivial damping term $\dot{x}f(x)$ in equations (1).
The van der Pol oscillator of equation $\ddot{x}-\epsilon (1-x^2)\dot{x}+x=0$,
is probably the simplest example of a system with one stable 
limit cycle: a fixed point at the origin and an unstable closed orbit
for $\epsilon<0$ and an attractive cycle for $\epsilon>0$.
A natural extension of the van der Pol system to the representation
of two symmetric coupled oscillators is:
\begin{eqnarray}
\ddot{x}_1-\epsilon (1-(x_1^2+x_2^2))\dot{x}_1+x_1+\beta x_2 & = & 0 \nonumber \\
\ddot{x}_2-\epsilon (1-(x_1^2+x_2^2))\dot{x}_2+x_2+\beta x_1 & = & 0
\end{eqnarray}
The small displacements near
$x_1=x_2=\dot{x}_1=\dot{x}_2=0$ are damped to zero 
when $\epsilon<0$ and give sustained oscillations when $\epsilon>0$.
The birth of a stable limit cycle is then governed by
the parameter $\epsilon$ and our interest rests in this regime ($\epsilon>0$).
Other properties of (2):\newline
(i) If $\beta=0$, equations (2) present an $O(2)$ symmetry.
If the complex variable  $z=x_1+ix_2$ is defined, system (2) becomes:
$\ddot{z}-\epsilon (1-|z|^2)\dot{z}+z=0$,
which has the symmetries: $z\rightarrow e^{i\phi} z$ and $z\rightarrow\bar{z}$.
A stable solution is $z=e^{i\phi}r(t)$ with $\phi$ constant and $r(t)$
solution of: $\ddot{r}-\epsilon (1-r^2)\dot{r}+r=0$.
Its representation point
is a straight line through the origin in the $(x_1,x_2)$ plane
at constant angle $\phi$, $r(t)$ oscillating along this line.
The periodic solution $z=e^{\pm it}$ is unstable.\newline
(ii) If $\beta\neq 0$ the phase symmetry is destroyed, although a $Z(2)$ symmetry
$z\rightarrow \pm i\bar{z}$ remains. The equations (2) become:
\begin{equation}
\ddot{z}-\epsilon (1-|z|^2)\dot{z}+z+i\beta\bar{z}=0
\end{equation}
Now there are two oscillating solutions: the symmetric mode $\Theta_1\equiv (x_1=x_2)$
given by $z_1=e^{i\pi/4}r_1(t)$ and the antisymmetric mode $\Theta_2\equiv (x_1=-x_2)$
given by $z_2=e^{-i\pi/4}r_2(t)$, where $r_1(t)$ and $r_2(t)$ verify:
$\ddot{r}_i-\epsilon (1-r_i^2)\dot{r}_i+(1-(-1)^i\beta)r_i =0$ (i=1,2).

This system presents stable oscillations when $|\beta|<1$ and diverges
to infinity when  $|\beta|>1$ (except for initial conditions $r=\dot{r}=0$).
One finds a parameter value $\beta_{\epsilon}>0$ such that:
if $0<\beta<\beta_{\epsilon}$ then $\Theta_1$
is stable and $\Theta_2$ unstable and, 
if $-\beta_{\epsilon}<\beta<0$, $\Theta_1$
is unstable and $\Theta_2$ stable. Also, if $\beta_{\epsilon}<|\beta|<1$
the two modes $\Theta_1$ and $\Theta_2$ are linearly stable.
Let us remark that
this simple model brings all the information we are looking for.
The range of parameters $\epsilon>0$ would modelize the situations
of stable limit cycle oscillations observed experimentally for
for $Re>Re_c$, where $Re_c$ is the Reynolds number
 at the onset of vortex shedding.
 The parameter $\beta$ would represent for instance
the aspect ratio in the experiments
of Le Gal and collaborators \cite{legal}.\par

In model equation (2) the transition from $\Theta_1$ 
to $\Theta_2$ stable oscillation occurs at $\beta=0$. 
In this case the coupling is lost and the system 
becomes degenerate at transition
(the phase difference between the two oscillators is arbritary)
presenting an infinity of stable oscillating states.
In order to remove this degeneracy,
we need to have more than one coupling parameter.
This means that the dimension of parameter space for a transition
between two modes of oscillation should be greater than one:
the unfolding of this transition should be controlled
by two parameters at least. Thus,
a more general and 'robust' scheme of transition 
from mode $\Theta_1$ to mode $\Theta_2$,
and viceversa, is achieved by introducing another
phase symmetry breaking term (proportional to $\gamma$)
in the dissipative force of equations (2):
\begin{eqnarray}
\ddot{x}_1-\epsilon (1-x_1^2-(1+\gamma )x_2^2))\dot{x}_1+x_1+\beta x_2 & = & 0 \nonumber \\
\ddot{x}_2-\epsilon (1-x_2^2-(1+\gamma )x_1^2))\dot{x}_2+x_2+\beta x_1 & = & 0
\end{eqnarray}
where $\gamma$ and $\beta$ are the coupling constants.
Symmetries $(x_1,x_2)\leftrightarrow (x_2,x_1)$ 
and $(x_1,x_2,\beta)\leftrightarrow (x_2,-x_1,-\beta)$ are preserved. \newline
We have numerically studied the solutions of this system 
in two different regimes and found the following
results for $\gamma$ and $\beta$ near zero : \newline
(a) When $|\beta|\ll|\gamma|$ there are four oscillatory states: the pure symmetric
mode $\Theta_1\equiv (x_1=x_2)$, the pure antisymmetric mode
$\Theta_2\equiv (x_1=-x_2)$ and two new mixed modes $\Theta_{12}\equiv (x_1,x_2)$
and $\Theta_{21}\equiv (x_2,x_1)$ intermediate between $\Theta_1$ and $\Theta_2$.
If $\gamma>0$ the mixed modes are stable and the pure modes unstable.
If $\gamma<0$ the mixed modes are unstable and the pure modes stable. \newline
(b) When $|\beta|\gg|\gamma|$ equations (4) tend to equations (2)
(the perturbation introduced by $\gamma$ can be neglected in front of $\beta$),
the mixed  modes $\Theta_{12}$ and $\Theta_{21}$ collide and disappear, and
the pure mode $\Theta_{1}$ and $\Theta_{2}$ remain.

Let us explain in more detail the two different scenarios (Fig. 2) that
can be found for the transition between the pure modes $\Theta_1$ and $\Theta_2$
when $\gamma$ is fixed and $\beta$ is varied ($\epsilon$ is kept
constant and of order $1$, but the results are not sensitive to its
specific value): \newline
\underline{\underline{Scenario I}}; \underline{$\gamma<0$} (Fig. 2a):\newline
(I$_1$) $\beta<-c(\epsilon)|\gamma|$ ($c(\epsilon)$ positive constant,
depending on $\epsilon$ and of order $1$ for $\epsilon$ of order $1$):
 $\Theta_{1}$ is unstable
 and $\Theta_{2}$ stable. No mixed modes. \newline
(I$_2$) $-c(\epsilon)|\gamma|<\beta<c(\epsilon)|\gamma|$: 
the two unstable mixed modes $\Theta_{12}$ and $\Theta_{21}$
grow from $\Theta_{1}$ for $\beta=-c(\epsilon)|\gamma|$. In this regime the two pure
modes are stable. Depending on initial conditions the system oscillates
in the symmetric or in the antisymmetric mode. 
When $\beta\rightarrow c(\epsilon)|\gamma|$
the two mixed modes approach $\Theta_{2}$, collide with it for $\beta=c(\epsilon)|\gamma|$
making $\Theta_2$ linearly unstable.
It transfers the stability from $\Theta_{2}$ to $\Theta_{1}$. \newline
(I$_3$) $\beta>c(\epsilon)|\gamma|$: $\Theta_{1}$ is stable 
and $\Theta_{2}$ unstable. No mixed modes. \newline
Summarizing: there is a range of parameters $I_2$ where  $\Theta_{1}$ and $\Theta_{2}$
are both stable, and each mode ( $\Theta_{1}$ or  $\Theta_{2}$) loses
its stability by a supercritical bifurcation on the edges of $I_2$.\newline
\underline{\underline{Scenario {II}}}; \underline{$\gamma>0$} (Fig. 2b):\newline
({II}$_1$) $\beta<-c(\epsilon)|\gamma|$: $\Theta_{1}$ 
is unstable and $\Theta_{2}$ stable. No mixed modes. \newline
({II}$_2$) $-c(\epsilon)|\gamma|<\beta<c(\epsilon)|\gamma|$:
 the two mixed modes $\Theta_{12}$ and $\Theta_{21}$
bifurcate from $\Theta_{2}$ for $\beta= -c(\epsilon)|\gamma|$. These are stable
 (which makes the difference with scenario I). In this ${II}_2$ regime the two pure
modes are unstable and the system will decay in one of the two mixed modes
according to the initial conditions. When $\beta\rightarrow c(\epsilon)|\gamma|$
the mixed modes approach $\Theta_{1}$ and collide with it for $\beta=c(\epsilon) |\gamma|$.
It transfers the stability from the mixed modes to $\Theta_{1}$. \newline
({II}$_3$) $\beta>c(\epsilon)|\gamma|$: $\Theta_{1}$ is stable
 and $\Theta_{2}$ unstable. No mixed modes.\newline
Sumarizing, there is a range of parameters ${II}_2$ where the two mixed modes
are stable, and collide with  $\Theta_{1}$ or  $\Theta_{2}$ on the edge of
${II}_2$ to exchange stability.

A derivation of the dynamics of equations (4) can be obtained in the 
formalism of a slow phase dynamics \cite{manneville}.
(A different calculation can be found in reference \cite{aronson}
where two nonlinear oscillators with diffusive coupling,
not the one we consider, 
are studied in the vicinity of a Hopf (Poincar\'{e}-Andronov) bifurcation).
When $\beta=\gamma=0$,
the set of equations (4) present phase symmetry ($\phi$) 
and temporal translation symmetry ($\psi$),
 and when $\beta$ or $\gamma$ are different
from zero the phase symmetry is broken. For small $\beta$ or $\gamma$
(of the same order of magnitude) the general solution of equations (4)  can be written:
\begin{eqnarray*}
x_1 & = & r_0(t+\psi)\cos\phi+\delta x_1 \\
x_2 & = & r_0(t+\psi)\sin\phi+\delta x_2 
\end{eqnarray*}
where $r_0(t)$ is the periodic non zero solution of the van der Pol equation: 
$\ddot{r}_0-\epsilon (1-r_0^2)\dot{r}_0+r_0=0$,
 $\psi$ and $\phi$ follow a slow dynamics
($\frac{\dot{\psi}}{\psi},\frac{\dot{\phi}}{\phi}\ll \frac{\dot{r_0}}{r_0}$),
and $\delta x_1$, $\delta x_2$, $\dot{\psi}$, $\dot{\phi}$ 
are small and of order $(\gamma,\beta)$.
Linearising equations (4) to order $(\gamma,\beta)$ we obtain a set of coupled equations 
to be solved for $\delta x_1$ and $\delta x_2$. These equations are written 
in matrix notation to make their structure more transparent:
\begin{equation}
{\cal L}\left(\begin{array}{c} \delta x_1 \\ \delta x_2 \end{array}\right)
= \left(\begin{array}{cc}
f(r_0)\sin\phi & g(r_0)\cos\phi \\
-f(r_0)\cos\phi & g(r_0)\sin\phi \end{array}\right)
\left(\begin{array}{c} \dot{\phi}  \\ \dot{\psi} \end{array}\right)- 
r_0[\beta+\frac{\gamma}{2}h(r_0,\phi)]
\left(\begin{array}{c} \sin\phi \\ \cos\phi \end{array}\right)
\end{equation}
where
\begin{eqnarray*}
{\cal L} & = & 
\left(\begin{array}{cc}
{\cal F}_t & 0 \\
0 & {\cal F}_t
\end{array}\right) + h(r_0,\phi)
\left(\begin{array}{cc}
\cot\phi & 1 \\
1 & \tan\phi \end{array}\right) \\
{\cal F}_t & = & \partial_{tt}-\epsilon (1-r_0^ 2)\partial_t+1 \\
f(r_0) & = & 2\dot{r}_0-\epsilon(1-r_0^2)r_0 \\
g(r_0) & = & -[2\ddot{r}_0-\epsilon(1-r_0^2)\dot{r}_0] \\
h(r_0,\phi) & = & \epsilon r_0\dot{r}_0\sin(2\phi)
\end{eqnarray*}
The relevant solution of equation (5) is made of periodic functions of time,
with the same period $T$ as $r_0(t)$. This excludes functions with a secular growth
and leads to a solvability condition that will ultimately become an equation
of evolution for $\phi(t)$. To write this solvability condition, one needs 
to define first an inner product of functions of time with period $T$ as:
$<\vec{\theta}|\vec{\sigma}>\equiv \int_0^{T} (\theta_1\sigma_1+\theta_2\sigma_2)dt$,
($\vec{\theta}=(\theta_1,\theta_2)$ is written as a two component vector).
One notices now that the linear operator $\cal L$ has a nonempty kernel:
\begin{eqnarray*}
{\cal L}\vec{\omega}=0 & \Rightarrow & 
\vec{\omega}_a=r_0\left(\begin{array}{c} \sin\phi \\ -\cos\phi \end{array}\right)\;\; ,
\;\; \vec{\omega}_b=\dot{r}_0\left(\begin{array}{c} \cos\phi \\ 
\sin\phi \end{array}\right)
\end{eqnarray*}
Because of this non empty kernel, the equation (5) has no solution in general
that is periodic with period $T$. To have such a solution, the right 
hand side $\vec{\varphi}$
of this equation must be orthogonal to the 
kernel of the adjoint operator ${\cal L}^+$,
made of two functions, $\vec{\chi}_i$ ($i=a,b$), of $t$ that 
are solution of the formal equation  ${\cal L}^+\vec{\chi}=0$.
The solvability condition is then that the two inner products 
$<\vec{\chi}_i|\vec{\varphi}>$ ($i=a,b$) are zero. 
The operator ${\cal L}^+$ can be written explicitely as:
\begin{eqnarray*}
{\cal L}^+ & = & 
\left(\begin{array}{cc}
{\cal F}_t^+ & 0 \\
0 & {\cal F}_t^+
\end{array}\right) + h(r_0,\phi)
\left(\begin{array}{cc}
\cot\phi & 1 \\
1 & \tan\phi \end{array}\right) 
\end{eqnarray*}
where 
${\cal F}_t^+ = \partial_{tt}+\epsilon\partial_t(1-r_0^2)+1$.
Since the two left vectors $\vec{\chi}_{a,b}$, once multiplied with
the inner product, $<|>$, with the left side of equation (5) give zero,
the same product with the right side of (5) should give zero as well.
This gives two coupled equations for $\dot{\phi}$ and $\dot{\psi}$:
\begin{equation}
\left(\begin{array}{cc}
h_{a1} & h_{a2} \\
h_{b1} & h_{b2}
\end{array}\right)
\left(\begin{array}{c}
\dot{\phi} \\ \dot{\psi}
\end{array}\right) = \beta
\left(\begin{array}{c}
m_a \\ m_b
\end{array}\right) + \gamma
\left(\begin{array}{c}
n_a \\ n_b
\end{array}\right)
\end{equation} 
where $h_{i1}$, $h_{i2}$, $m_i$, $n_i$ $(i=a,b)$ are functions of
$\epsilon$ and $\phi$ after the time integration coming from the 
scalar product:
\begin{eqnarray*}
h_{i1} & = & \int_0^Tf(r_0)[\sin\phi\;\chi_{i1}-\cos\phi\;\chi_{i2}]dt \\
h_{i2} & = & \int_0^Tg(r_0)[\cos\phi\;\chi_{i1}+\sin\phi\;\chi_{i2}]dt \\
m_i & = & \int_0^Tr_0[\sin\phi\;\chi_{i1}+\cos\phi\;\chi_{i2}]dt \\
n_i & = & \frac{1}{2}\int_0^Tr_0h(r_0,\phi)[\sin\phi\;\chi_{i1}+\cos\phi\;\chi_{i2}]dt
\end{eqnarray*}
The $\phi$-dependance of the vectors in the kernel of ${\cal L}^+$ 
can be factored out by noticing that these vectors can have the following
$\phi$-dependance:
\begin{displaymath}
\vec{\chi}_a=h_a(t)\left(\begin{array}{c} -\sin\phi \\ \cos\phi \end{array}\right)\;\; ,
\;\; \vec{\chi}_b=h_b(t)\left(\begin{array}{c} \cos\phi \\ 
\sin\phi \end{array}\right)
\end{displaymath}
From this the two functions $h_{a,b}(t)$ are the non trivial (=non zero)
 solutions of period $T$ of the two linear homegeneous equations:
\begin{eqnarray*}
{\cal F}_t^+[h_a(t)]=0 & \Rightarrow & 
[\partial_{tt}+\epsilon\partial_t(1-r_0^2)+1]h_a(t)=0 \\
\; [{\cal F}_t+2\epsilon (1-r_0^2)\partial_t][h_b(t)]=0 & \Rightarrow & 
[\partial_{tt}+\epsilon (1-r_0^2)\partial_t+1]h_b(t)=0 
\end{eqnarray*}
Then the equation for $\dot{\phi}$ is simplified to: 
\begin{eqnarray}
k_a(\epsilon)\dot{\phi} & = & l_a(\epsilon)\cos(2\phi)\beta +
 s_a(\epsilon)\sin(4\phi)\gamma 
\end{eqnarray}
where $k_{a}$, $l_{a}$ and $s_{a}$ are functions of $\epsilon$ only that
are proportional to various scalar product of functions 
on the equation (6) with $h_{a}(t)$. Thus,
$k_a(\epsilon)=-\int_0^Tf(r_0)h_a(t)dt$ ;
$l_{a}(\epsilon)=\int_0^Tr_0h_{a,b}(t)dt$ ; and 
$s_{a}(\epsilon)=\frac{\epsilon}{4}\int_0^Tr_0^2\dot{r}_0h_{a,b}(t)dt$.

The equation (7) presents, as $\beta$ and $\gamma$ vary, the same bifurcations
as the one found numerically for the original set of equations (4). To show this,
let us define $\beta'=\beta\frac{l_a(\epsilon)}{k_a(\epsilon)}$ and
$\gamma'=2\gamma\frac{s_a(\epsilon)}{k_a(\epsilon)}$, that will be considered now 
as the bifurcation parameters (the quantities $\frac{l_a(\epsilon)}{k_a(\epsilon)}$
and $\frac{s_a(\epsilon)}{k_a(\epsilon)}$ are constants of order $1$ at a fixed
finite value of $\epsilon$, and so can be get rid off by scaling). The fixed points
of (7) are roots (in $\phi$) of:
\begin{equation}
\beta'\cos(2\phi)+\frac{\gamma'}{2}\sin(4\phi) = 0
\end{equation}
or of $\cos(2\phi)=0$ or $\beta'+\gamma'\sin(2\phi)=0$.\newline
If $|\beta'|>|\gamma'|$, this corresponds to scenario $I_{1,3}$ and $II_{1,3}$.
The only steady states are at the zeroes of $\cos(2\phi)$, that are at 
$\phi=\frac{\pi}{4}$ and $\phi=-\frac{\pi}{4}$, with one stable and the other unstable,
depending of the sign of $\beta'$ (and consequently of $\beta$) in agreement with
what was found numerically. If $|\beta'|<|\gamma'|$, they are two more fixed 
points, that are  $\frac{1}{2}\sin^{-1}(-\frac{\beta'}{\gamma'})$ and 
$\frac{\pi}{2}-\frac{1}{2}\sin^{-1}(-\frac{\beta'}{\gamma'})$. They correspond 
to the mixed modes and, as $\beta'$ goes for instance from $-\gamma'$ to $\gamma'$
(if $\gamma'>0$), one finds the same bifurcation structure as found for the original
equations (4), as explained formerly under the heading "Scenario $II$" (Fig. 2).

In this rapid communication, we have presented a simple model for systems made of two
symmetric coupled self-oscillators \cite{villermeux}. This might be a theory for one of
the most studied instabilities in real fluid mechanics, the periodic
release of vortices in the wake of cylinders, a phenomenon
studied experimental and theoretically long ago by 
H. B\'{e}nard and T. von Karman \cite{benard,karman} and his collaborators.
The connection of the present work
with the B\'{e}nard-von Karman phenomenon could be as follows. 
Our idea is that the wake is created by two symmetrically coupled self-oscillators,
one on each side of the cylinder. We have shown that, depending 
on the coupling, these two systems may either oscillate in phase
or out of phase (as in the B\'{e}nard-von Karman wake in a normal viscous fluid).
Moreover, the transition from one of these two states to the other is 
realized by two different scenarios depending of the parameters.
This might describe recent experiments by Le Gal
and collaborators \cite{legal}, who observe this transition
 when the flow around the cylinder
is more and more constrained by plates perpendicular to the axis of this
cylinder.

{\bf Acknowledgements:} We thank P. Le Gal for showing us some experimental results.
Y. P. would like to thank E. Villermaux for sharing many insights on this
topic of symmetrically coupled self-oscillators.
R. L-R. also thanks S. Rica for useful discussions and 
the European Community by a research grant.

\newpage
\begin{center} {\bf Figure Captions} \end{center}

{\bf Fig 1.} A graphical representation of the two oscillation modes: (a) the symmetric
vortex street $\Theta_1$ and (b) the antisymmetric one $\Theta_2$.

{\bf Fig 2.} Nonlinear transition between the two oscillation modes 
$(\Theta_1,\Theta_2)$:
 (a) in scenario I ($\gamma<0$) 
the two intermediary mixed modes, $\Theta_{12}$ and $\Theta_{21}$, are unstable and
(b) in scenario II ($\gamma>0$) these mixed modes are stable. (c) Another 
representation of scenarios I and II (inspired from figure 3 in reference \cite{aronson}).

\end{document}